\documentclass[namedreferences]{solarphysics}
\usepackage[hyperref,optionalrh,solaromanenum]{spr-sola-addons} 
\usepackage{graphicx}                    
\usepackage{color}                       
\usepackage{breakurl}                         

\usepackage{comment}

\begin{document}

\begin{article}

\begin{opening}

\title{Solar Energetic Particle Ground-Level Enhancements and the Solar Cycle}

%
\author[addressref={reading},corref,email={m.j.owens@reading.ac.uk}]{\inits{M.J.}\fnm{Mathew J.}~\lnm{Owens}\orcid{0000-0003-2061-2453}}

\author[addressref={reading}
]{\inits{L.A.}\fnm{Luke A.}~\lnm{Barnard}\orcid{0000-0001-9876-4612}}

\author[addressref={UQ}
]{\inits{B.}\fnm{Benjamin~J.\,S.}~\lnm{Pope}\orcid{0000-0003-2595-9114}}

\author[addressref={reading},
]{\inits{M.}\fnm{Mike}~\lnm{Lockwood}\orcid{0000-0002-7397-2172}}

\author[addressref={Oulu}
]{\inits{I}\fnm{Ilya}~\lnm{Usoskin}\orcid{numbers}}

\author[addressref={Helsinki}
]{\inits{E}\fnm{Eleanna}~\lnm{Asvestari}\orcid{numbers}}

%
\runningauthor{M.J. Owens et al.}
\runningtitle{Ground-Level Enhancements and the Solar Cycle}

\address[id={reading}]{Department of Meteorology, University of Reading, Earley Gate, PO Box 243, Reading RG6 6BB, UK}

\address[id={UQ}]{School of Mathematics and Physics, University of Queensland, St Lucia, QLD 4072, Australia}

\address[id={Oulu}]{Space Physics and Astronomy Research Unit and Sodankyla Geophysical Observatory, University of Oulu, 90014 Oulu, Finland }

\address[id={Helsinki}]{Department of Physics, University of Helsinki, Finland}

\begin{abstract}
Severe geomagnetic storms appear to be ordered by the solar cycle in a number of ways. They occur more frequently close to solar maximum and declining phase, are more common in larger solar cycles and show different patterns of occurrence in odd- and even-numbered solar cycles. Our knowledge of the most extreme space weather events, however, comes from the spikes in cosmogenic-isotope ($^{14}$C, $^{10}$Be and $^{36}$Cl) records that are attributed to significantly larger solar energetic particle (SEP) events than have been observed during the space age. Despite both storms and SEPs being driven by solar eruptive phenomena, the event-by-event correspondence between extreme storms and extreme SEPs is low. Thus it should not be assumed a priori that the solar cycle patterns found for storms also hold for SEPs and the cosmogenic-isotope events. In this study we investigate the solar cycle trends in the timing and magnitude of the 67 SEP ground-level enhancements (GLEs) recorded by neutron monitors since the mid 1950s. Using a number of models of GLE occurrence probability, we show that GLEs are around a factor four more likely around solar maximum than around solar minimum, and that they preferentially occur earlier in even-numbered solar cycles than in odd-numbered cycles. There are insufficient data to conclusively determine whether larger solar cycles produce more GLEs. Implications for putative space-weather events in the cosmogenic-isotope records are discussed. We find that GLEs tend to cluster within a few tens of days, likely due to particularly productive individual active regions, and with approximately 11-year separations, owing to the solar cycle ordering. But these timescales do not explain cosmogenic-isotope spikes which require multiple extreme SEP events over consecutive years.
\end{abstract}

%

\end{opening}

%

\section{Introduction}
\label{sec:intro}

Severe space weather, in the form of both extreme geomagnetic storms \citep{gosling_solar_1993,richardson_sources_2002, richardson_near-earth_2010} and severe solar energetic particle (SEP) events \citep{reames_two_2013,desai_large_2016}, is associated with solar eruptive phenomena such as coronal mass ejections (CMEs). Determining how the occurrence and magnitude of these rapid, localised solar eruptions varies with the global solar activity cycle is central to both long-lead time space-weather forecasting and understanding of the processes by which the solar atmosphere reconfigures and sheds excess magnetic flux and helicity \citep{low_coronal_2001,owens_coronal_2006,lynch_solar_2005}.

While direct measurements of the solar wind are limited to the last 60 years or so \citep{cliver_solar_1994}, ground-based magnetometer records measure the subsequent disturbance to the terrestrial system and extend back around 170 years \citep{svalgaard_idv_2005,lockwood_reconstruction_2013-1,owens_near-earth_2016-1}. Less direct proxies for historic solar activity also exist. Four centuries of telescopic sunspot observations \citep{clette_new_2016} allow long-term trends in global solar magnetism to be inferred, though do not record individual space-weather events. Relating solar activity on short time scales (i.e. space weather) to that on much longer time scales (i.e. space climate, such as the solar cycle and grand minima and maxima, see \cite{mursula_introduction_2007}) would be practically advantageous, as there is predictability on decadal timescales. While advanced predictions of the amplitude of Solar Cycles 24 and 25 spanned a large range of values \citep{pesnell_lessons_2020, nandy_progress_2021}, the fact that the solar cycle is nominally between 10 and 12 years in length \citep[e.g.][]{van_driel-gesztelyi_solar_2020} means that the time of solar maximum can be predicted to within a year or two with reasonable confidence. Thus useful information for long-term planning can be gained if the occurrence of hazardous space weather events follows the phase of the solar cycle, at least in a statistical sense \citep[e.g. ][]{kilpua_statistical_2015, vennerstrom_extreme_2016,chapman_quantifying_2020}. Recently, probabilistic models were used to show that the occurrence of extreme geomagnetic storms does follow the solar cycle, at least up to largest magnitudes that can reasonably be tested, around 1-in-20 year level events \citep{owens_extreme_2021}. 

On longer time scales, cosmogenic radionuclides, particularly $^{14}$C and $^{10}$Be in tree trunks and ice cores, provide information about long-term trends in global solar magnetic field strength \citep{usoskin_history_2017}. The flux of galactic cosmic rays (GCRs) reaching Earth is modulated by the strength of the heliospheric magnetic field (HMF). Thus the production rate of radionuclides, such as $^{14}$C and $^{10}$Be resulting from GCR-induced spallation of atmospheric constituents, can be used to infer the HMF \citep[e.g.][]{caballero-lopez_heliospheric_2004}. As the deposition and sequestration times for cosmogenic radionuclides from the atmosphere to trees and ice is a multi-year process, they are typically used to infer long-term variations, such as grand minima and maxima of solar activity \citep[e.g.][]{solanki_unusual_2004,steinhilber_9400_2012, usoskin_history_2017,owens_near-earth_2016} and, more recently, the timing and amplitude of individual solar cycles \citep{usoskin_solar_2021,brehm_eleven-year_2021}. 

Within the last decade, more rapid, short-term large-amplitude variations in $^{14}$C, $^{10}$Be and $^{36}$Cl  production rates have also been identified \citep{miyake_signature_2012,miyake_another_2013}, often referred to as `Miyake events'. The fact that such a sharp rise in $^{14}$C production is seen globally is taken as evidence that the driver is external to the terrestrial system, with astrophysical gamma-ray bursts \citep{hambaryan_galactic_2013}, supernova \citep{dee_supernovae_2017}, magnetars \citep{wang_consequences_2019}, and atmospheric deposition by comets \citep{liu_mysterious_2014} all proposed and disputed \citep{usoskin_carbon-14_2015}. But the now generally-accepted interpretation is that Miyake events are produced by the Sun \citep[e.g.][]{melott_causes_2012,usoskin_ad775_2013,cliver_solar_2014} and result from extreme SEP events. However, there are a number of points that do need to be addressed with this interpretation. Modelling suggests that the SEP events must be one--two orders of magnitude larger than any directly observed during the space age \citep{hudson_carrington_2021, cliver_size_2020, cliver_extreme_2022}. Observations of superflares on other stars suggests this may be possible \citep{okamoto_statistical_2021}, even for slowly-rotating stars like the Sun \citep{notsu_kepler_2019}. One of the Miyake events, ca.660 BC, also seem to require multi-year $^{14}$C production \citep{sakurai_prolonged_2020}. This would require temporal clustering on timescales of around a year for the most extreme (and therefore rare) SEP events, which will be investigated in this study. Further evidence of a common physical origin would be if both extreme SEPs (historically referred to as solar cosmic rays) and Miyake events show similar patterns of occurrence with respect to the solar cycle. However, we note that determining the phase of the solar cycle at which Miyake events occur is difficult, not least as the large SEP-driven perturbation to the $^{14}$C record obscures the underlying solar cycle trend \citep{usoskin_solar_2021}. 

In order to better understand both the physical origin of Miyake events and extreme space weather in general, it is instructive to determine the degree to which extreme SEP occurrence follows the solar cycle. Space-based observations of near-Earth SEPs are available back to 1967, though with a range of different instrument sensitivities and some data outages, particularly in the early 2000s \citep{barnard_survey_2011}. A longer and more contiguous record of the SEPs specifically relevant to Miyake events (i.e. producing effects detectable in the troposphere) can be obtained from the database of ground-level enhancements (GLEs). GLEs are SEP events with sufficiently hard spectra and large enough fluence to produce signatures detected by ground-level ionization chambers and neutron monitors \citep{forbush_three_1946,usoskin_revised_2020}. GLEs are of particular space-weather interest for the potential radiation effects on aviation \citep[e.g.][]{miroshnichenko_retrospective_2018}. Neutron monitors allow both the timing and magnitude of GLEs to be estimated. There are 68 known events in the continuous neutron monitor record since 1956 (see the International GLE database \url{https://gle.oulu.fi}). By eye, they do appear to approximately follow the solar cycle \citep{shea_summary_1990,barnard_what_2018}, but due to the low number of events, it is not clear whether this is by chance or due to an underlying tendency. Furthermore, many of the GLEs are very small increases above the background level and it is of particular interest whether the few strongest GLEs follow the solar cycle. 

Here we apply statistical methods to determine the correspondence between extreme GLE occurrence and the solar cycle. Section \ref{sec:data} describes the datasets used and where they may be obtained, as well as the association (or lack thereof) between extreme geomagnetic storms and GLEs. Section \ref{sec:models} outlines the probability models that are used to test the significance of GLE occurrence trends. Section \ref{sec:results} describes the results for each aspect of the solar cycle trends in GLEs and the clustering of GLEs in time. Finally, Section \ref{sec:discussion} discusses the implications of the results for space weather and interpretation of Miyake events.

\section{Data}
\label{sec:data}

\begin{figure}[!t]
 \centerline{\includegraphics[width=1\textwidth,clip=]{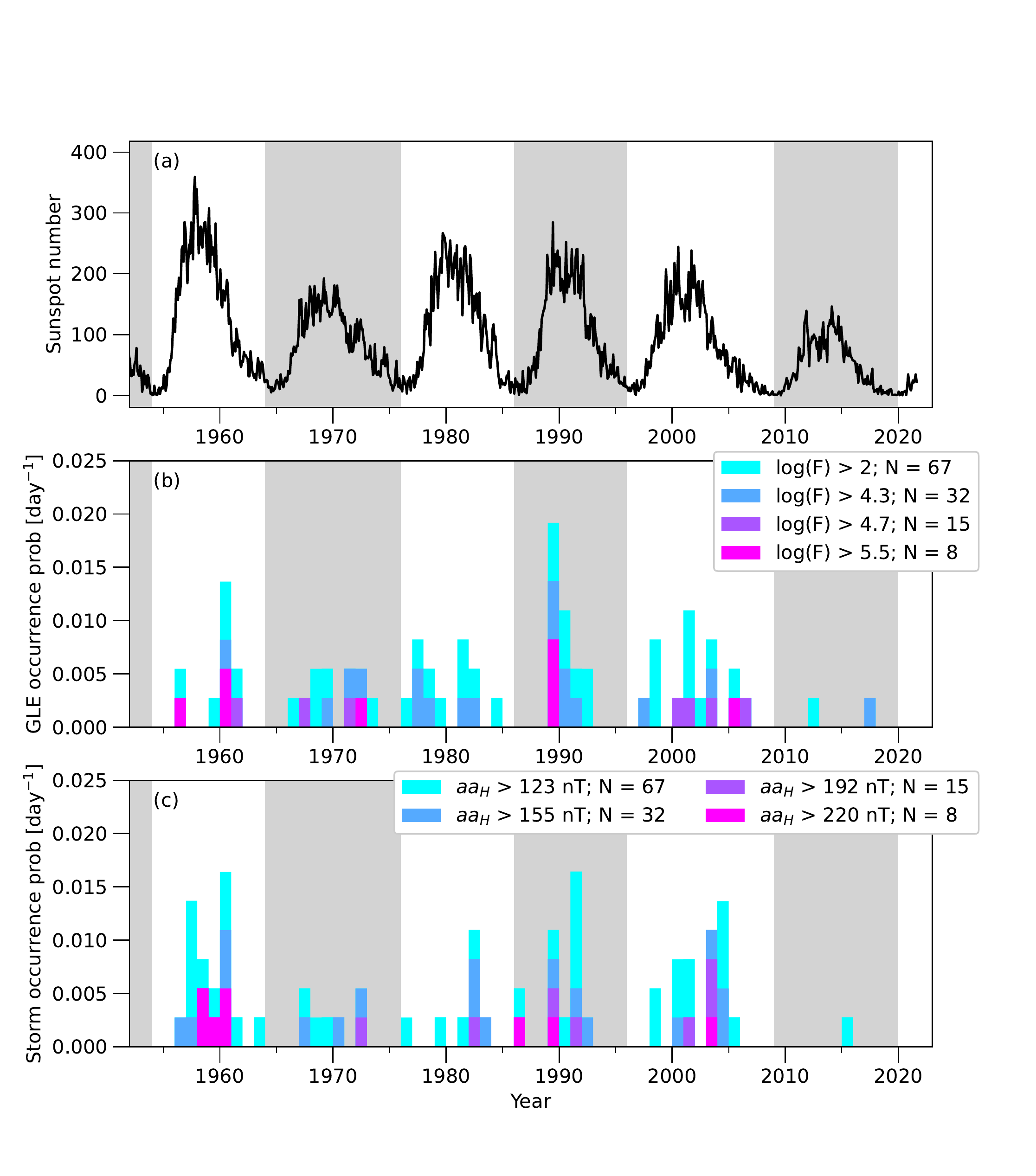}}
 \caption{Time series of (a) sunspot number, (b) GLE occurrence and (c) geomagnetic storm occurrence, as determined from the  aa$_\mathrm{H}$ dataset. White/grey shaded regions show odd/even solar cycles, respectively.}\label{fig:timeseries}
 \end{figure}
 
The first four recorded GLEs were measured using ionisation chambers \citep{forbush_three_1946} and, as such, there are no estimates of event magnitude. The timing of GLEs during the neutron monitor era, from the 1950s onwards \citep{stoker_igy_2009}, is provided by the GLE database \url{https://gle.oulu.fi/}) see \cite{asvestari_analysis_2017} for more detail. This database covers period from 1956 to early 2022 and includes 68 events. A number of smaller, `sub-GLE', events have been identified in the more recent data \citep{miroshnichenko_small_2020}, but these are not included in the present study so as to maintain a consistent event threshold throughout the 66-year interval considered. In order to estimate the GLE magnitude, we use the inferred spectral parameters to produce estimates of integrated fluence, $F$, above some rigidity threshold, $R$, using Equation 5 of \cite{usoskin_revised_2020}:
\begin{equation}
  F(> R) = F_0 \left( \frac{R}{1\, \textrm{GV}} \right) ^ \gamma  \exp \left( \frac{- R}{R_0} \right)
\end{equation}
where $F_0$ is a normalisation coefficient, $\gamma$ is the spectral index, and $R_0$ is the roll-off rigidity. Taking $R$ = 1 GV, and substituting for $J_0 = 4 \pi F_0$, where $J_0$ is the integral intensity, this simplifies to:

\begin{equation}
    F (> 1\, \textrm{GV})  = \frac{J_0}{4\pi} \exp \left( \frac{-1}{R_0} \right)
\end{equation}

The spectral parameters $J_0$ and $R_0$ are provided by \cite{usoskin_revised_2020} for 53 of the neutron-monitor era GLEs, giving $F(> 1\, \textrm{GV})$ values in the range $2 \times 10^3$ to $6 \times 10^6$ cm$^{-2}$. For the remainder of the study, $F(> 1\, \textrm{GV})$ [cm$^{-2}$] is simply referred to as $F$. The remaining 14 GLEs were deemed to be too small to enable reliable estimation of the spectrum. Here we use nominal values of $J_0 = 10^4$ and $R_0 = \infty$, which results in $F = 8 \times 10^2$ cm$^{-2}$, to enable the timing of these small events -- which is accurately known -- to be included in the study. These data are provided as part of the supplementary material to this paper.

From this dataset, we produce a GLE-day time series for a range of $F$ thresholds. Including all the events (i.e. $\log(F) > 2$) gives the full 67 events (one event is excluded, as described below). A threshold of $\log(F) > 4.3$ approximately halves the number of events (32), while $\log(F) > 4.7$ and $\log(F) > 5.5$ give 15 and 8 events, respectively. The results presented here are not particularly sensitive to the choice of these thresholds, as can be verified using the provided code and data.

To compare GLEs with geomganetic storms, we use the  aa$_\mathrm{H}$ index of geomagnetic activity \citep{lockwood_homogeneous_2018,lockwood_homogeneous_2018-1}, averaged to 1 day. The aa$_\mathrm{H}$ index is based on the same data as the classical aa index but has been corrected to allow for secular change in the geomagnetic field and its effect on station response to space weather activity. In addition, calibrations between stations that allow for time-of-day and time-of-year have been employed which generates a much more homogeneous and accurate time series.  Thresholds of  aa$_\mathrm{H}$ are applied to select the days associated with the largest storms. Using daily  aa$_\mathrm{H}$ thresholds of 123, 155, 192, and 220 nT gives 67, 32, 15, and 8 events, respectively, viz. the same statistics as for the GLEs. This enables direct comparison with the GLEs.

The monthly sunspot number, available from \url{www.sidc.be/silso/datafiles}, is used to determined the solar cycle phase and magnitude. The start and end of solar cycles are identified using the discontinuous change in the average latitude of the sunspots, as outlined in \citet{owens_solar_2011}. We further assume that Solar Cycle 25 began at the end of 2020. Using these solar-minimum times, the solar-cycle phase is computed as zero at the start of the cycle and increasing linearly to unity at the end. Thus not knowing the length of Cycle 25, we cannot compute the phase for the GLE 2021-10-28 and exclude it from the main study.

Figure \ref{fig:timeseries} summarises these datasets. Panel a shows the monthly sunspot number, with the grey-shaded panels showing even-numbered solar cycles. Panel b shows the annual average occurrence probability of GLEs for the four $F$ thresholds. As all years contain complete data coverage, this is the number of events per year divided by the number of days in the year (i.e. 0.027 indicates one event in that year). Panel c shows the occurrence probability for geomagnetic storms, using  aa$_\mathrm{H}$ thresholds chosen to give 67, 32, 15, and 8 events (i.e. the same number of total events as for GLEs at the four thresholds). From the time series alone, it is difficult to assess the degree to which GLEs follow the solar cycle or which GLE days correspond to geomagnetic storm days.

\subsection{GLE Association with Extreme Geomagnetic Storms}

 \begin{figure}[!t]
 \centerline{\includegraphics[width=1\textwidth,clip=]{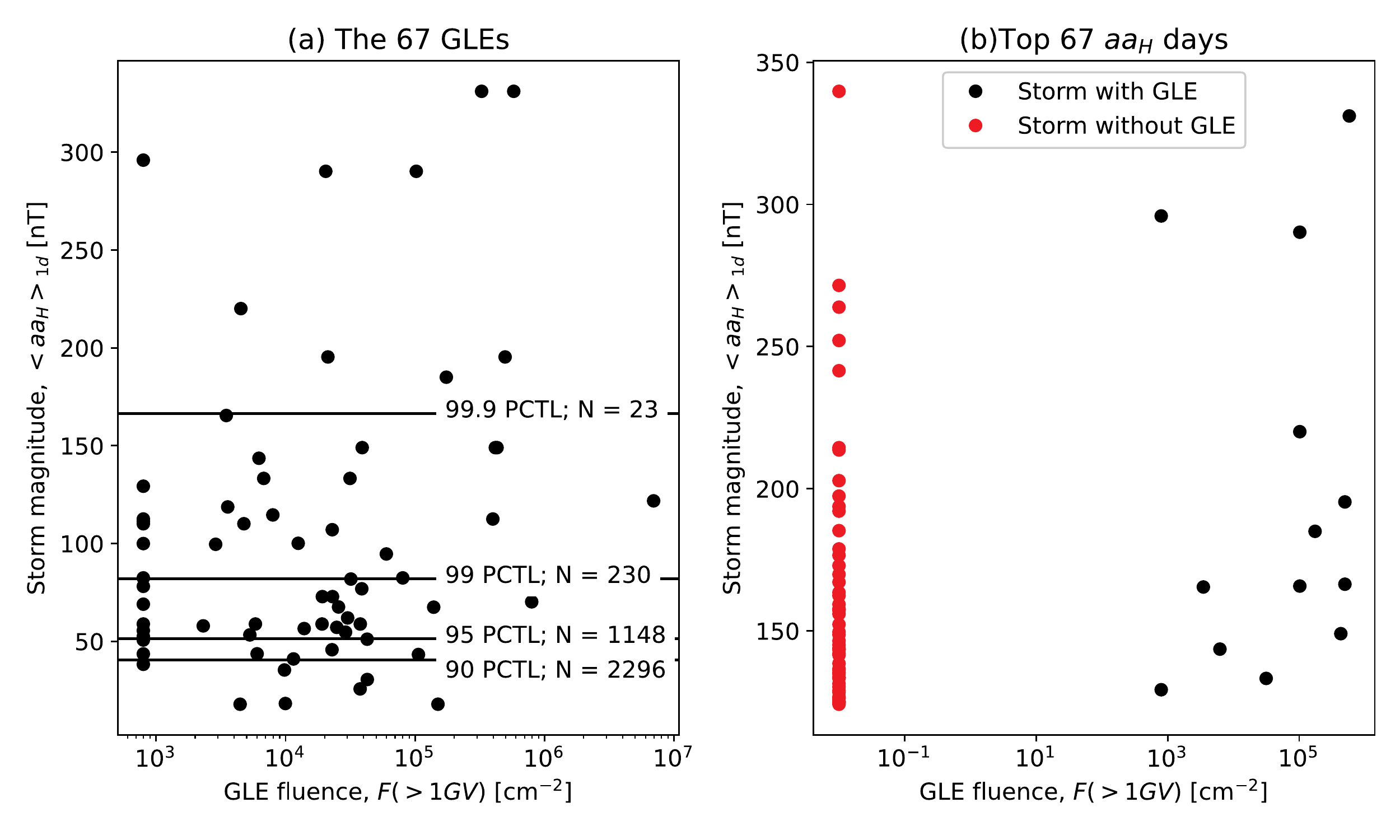}}
 \caption{The association between GLEs and geomagnetic storms (as measured by daily  aa$_\mathrm{H}$ index). (a) The maximum  aa$_\mathrm{H}$ within four days of each of the 67 GLEs as a function of GLE intensity. The black horizontal lines show percentile of  aa$_\mathrm{H}$ over the whole 1956-2020 period. 32 of the 68 GLEs are associated with storms above the 99th percentile (which produces 230 storm days). (b) Black: The 13 of the top 67  aa$_\mathrm{H}$ days in the 1956-2020 period for which a GLE was present within four days. Red: The 54 of the top 67 storms for which no GLE was present. An intensity of 0.01 was assigned for plotting purposes.}\label{fig:aaH}
 \end{figure}
 
Figure \ref{fig:aaH}a shows the largest  aa$_\mathrm{H}$ value within four days of the each GLE. There is no obvious correlation between GLE magnitude, as measured by $F$, and storm magnitude. More importantly, fewer than half the GLEs (32 of the 67) are associated with storms defined by the 99th percentile of  aa$_\mathrm{H}$, which produces 230 events in the period of study. The converse is shown in Figure \ref{fig:aaH}b: The top 67 geomagnetic storm days as a function of GLE fluence within 4 days. For 54 of these top 67 storms, there was no GLE. Thus there is little event-by-event correlation between storms and GLEs \citep[e.g.][]{nitta_what_2012}, and therefore we should not assume that the solar cycle trends found for geomagnetic storms apply to GLEs.

\subsection{GLE occurrence over the solar cycle}

 \begin{figure}[!t]
 \centerline{\includegraphics[width=1\textwidth,clip=]{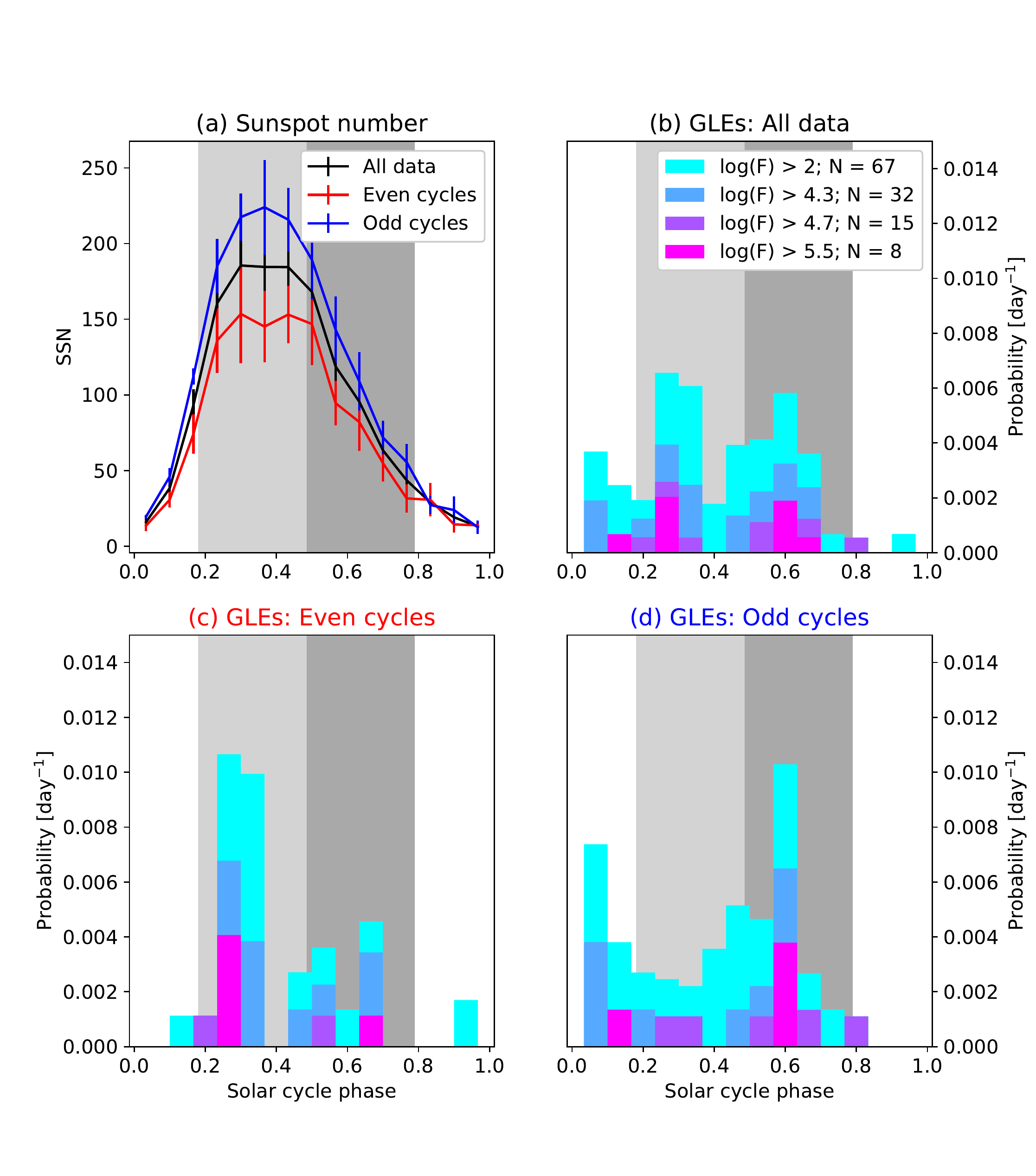}}
 \caption{Superposed epoch plots by solar cycle phase of (a) sunspot number and (b) GLE occurrence. Panels b and d show the GLE occurrence further divided in odd- and even-numbered solar cycles. Grey-shaded regions show the active phase of the solar cycle, identified with geomagnetic storms in \cite{owens_extreme_2021}. Light and dark grey further divide the active phases into early and late active phases.}\label{fig:SPE}
 \end{figure}

The time series shown in Figure \ref{fig:timeseries} hint at GLE occurrence following the solar cycle and there does appear to be some evidence for larger cycles producing more GLEs, with the two smaller sunspot cycles peaking in 1970 and 2015 containing fewer GLEs than the four larger cycles. This is investigated statistically in Section \ref{sec:amp}. 

In order to better visualise the ordering (or otherwise) of GLEs as a function of the solar cycle, Figure \ref{fig:SPE} shows a superposed epoch plot of GLE occurrence ordered by solar cycle phase. This normalises for variability in cycle length. For reference, grey-shaded panels show the active period identified for geomagnetic storms by \cite{owens_extreme_2021}: The active phase starts at a phase of 0.18 and ends at a phase of 0.79. The light- and dark-grey shading bisects the early and late active period. It can be seen that GLEs of all magnitudes appear to be more common in the active period than the quiet period. The statistical significance of this result will be tested in Section \ref{sec:phase}. 

Figures \ref{fig:SPE}c and d show the same analysis limited to even- and odd-numbered solar cycles, respectively. There is a prevalence of GLEs early in even-numbered cycles and late in odd-numbered cycles. Despite these GLEs being largely distinct events from extreme geomagnetic storms, they do appear to follow the same occurrence patterns: This same trend was reported for extreme geomagnetic storms \citep{owens_extreme_2021} and has been seen in more quiescent geomagnetic activity \citep{cliver_22year_1996} and solar wind conditions \citep{thomas_22-year_2013}. Of course, by bisecting the dataset into odd and even cycles, the number of GLEs considered in each category is reduced by around a half. Thus it cannot be ruled out that this apparent ordering is purely by chance, rather than due to an underlying physical trend. This is investigated in Section \ref{sec:parity}.

\section{Probability Models}
\label{sec:models}

 \begin{figure}[!t]
 \centerline{\includegraphics[width=1\textwidth,clip=]{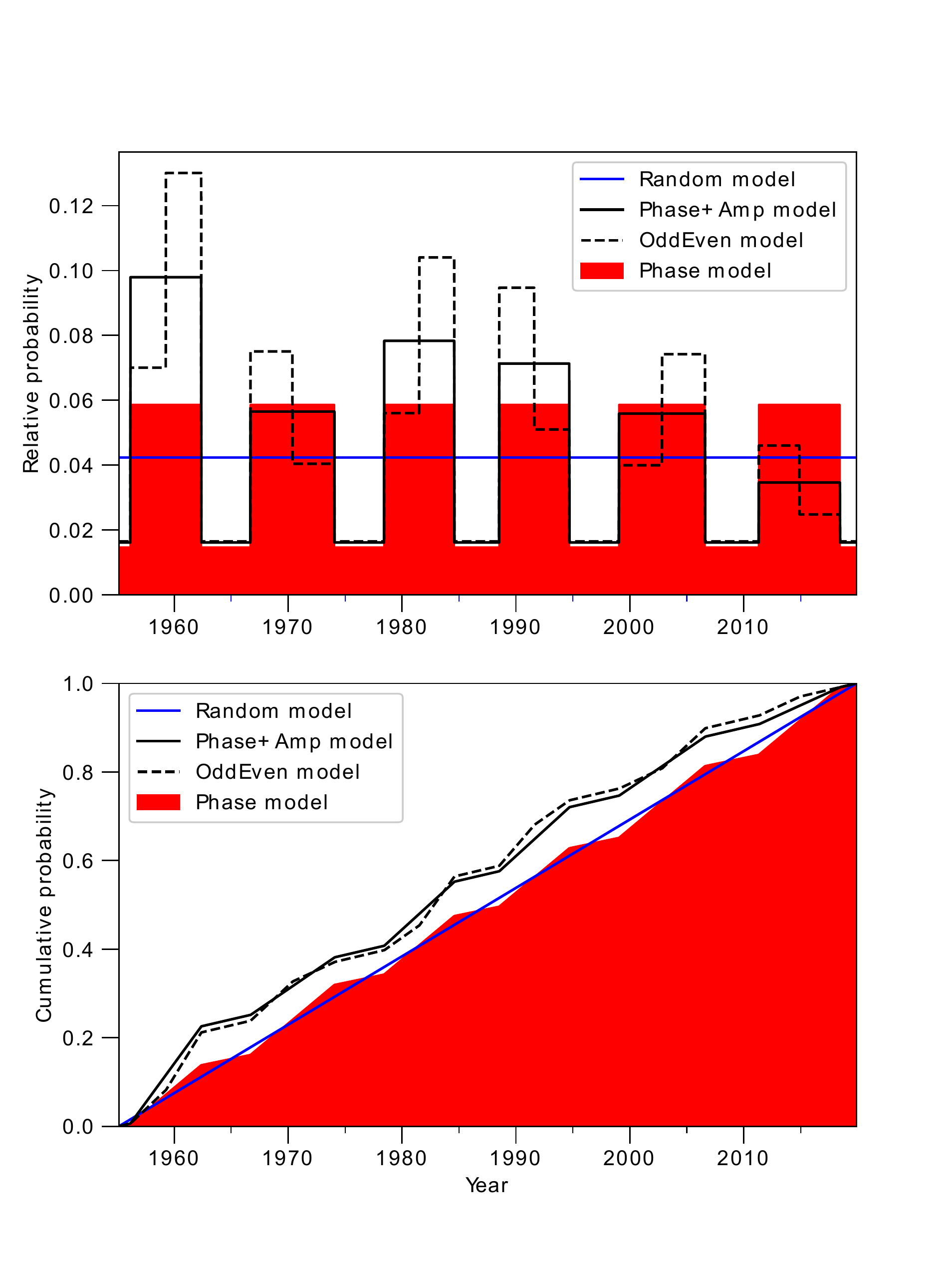}}
 \caption{Top: Models of GLE occurrence probability used to make inferences about the observations. Bottom: Cumulative distribution functions of the model probability, from which events are drawn. The Random model (blue) has a constant probability of a GLE at all times. The Phase model (red) has a factor four higher GLE probability during the active phase that the quiet phase. The Phase+Amp model further weights the active phase by the solar cycle amplitude (taken to be the cycle-average sunspot number). Finally, the OddEven model further adjusts the GLE occurrence probability up or down by a factor 0.3 during the early and late active phases based on the solar cycle parity (i.e. odd or even numbered).}\label{fig:models}
 \end{figure}

Following the same approach as \cite{owens_extreme_2021}, we test the apparent trends in GLE occurrence by comparing the observed occurrence with a number of different probability models. These are shown in the top panel of Figure \ref{fig:models}. Details are discussed below, but briefly, the models are:

\begin{itemize}
    \item \textbf{Random model}, shown by the solid blue line. In this model, the probability of GLE occurrence is equal at all times.  
    \item \textbf{Phase model}, shown by the red shaded area. In this model, the probability of a GLE is a factor four higher during the active phase of the solar cycle (phase between 0.18 and 0.79) than during the quiet phase (all other times). 
    \item \textbf{Phase+Amp model}, shown by the solid black line. This is the same as the Phase model, but the probability in the active phase of solar cycle $N$ is linearly related to the cycle amplitude, taken to be the average sunspot number over the cycle, $\langle SSN\rangle_N$. 
    \item \textbf{OddEven model}, shown by the black dashed line. This is the Phase+Amp model, with a reduction in probability by a factor $(1+d)$ early in the active phase of odd cycles and late in even cycles, and an increase in probability by a factor $(1-d)$ in the late active phases of odd cycles and early in even cycles. We use $d=0.3$. 
\end{itemize}

The coefficients in the Phase, Phase+Amp and OddEven models are chosen to approximately match the average values seen in Figure \ref{fig:SPE}, though we do not attempt a `best fit' of these values: the aim of the models is only to establish whether trends present in the data can or cannot be attributed to sampling issues. For the Phase+Amp and OddEven models, the probability in the active phase is further multiplied by a factor $\langle SSN\rangle_N/\langle SSN\rangle$, where $\langle SSN\rangle$ is the average SSN over the whole 1956\,--\,2021 interval, of 94.5.

From each probability model we construct the cumulative distribution functions, shown in the bottom panel of Figure \ref{fig:models}b. From these, random events are drawn to produce time series of event occurrence consistent with the probability models.

\section{Results}
\label{sec:results}

\subsection{Solar Cycle Phase}
\label{sec:phase}

 \begin{figure}[!t]
 \centerline{\includegraphics[width=0.5\textwidth,clip=]{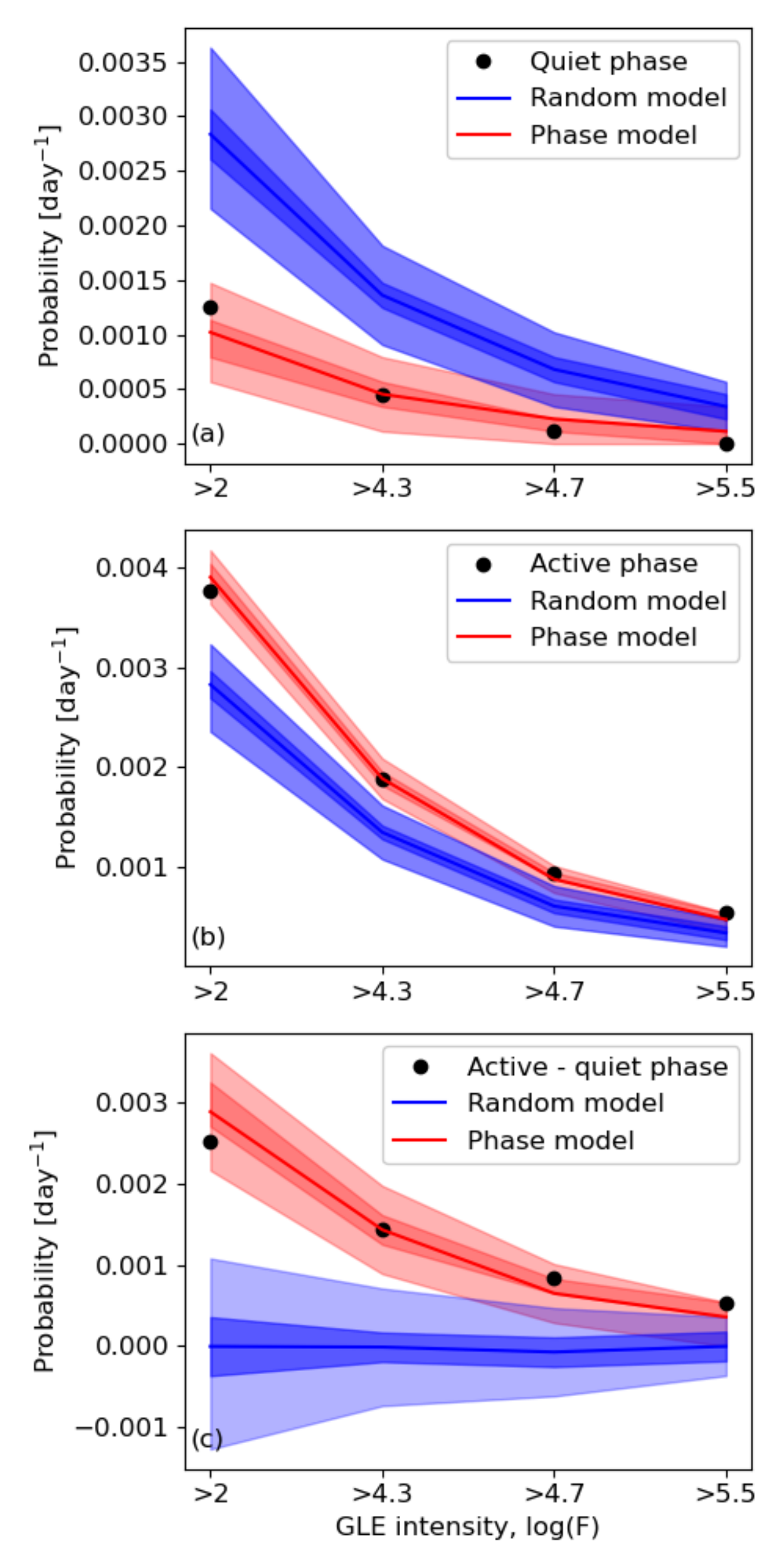}}
 \caption{GLE occurrence probability with increasing intensity threshold (a) in the quiet phase, (b) in the active phase, and (c) the difference between the active and quiet phases. Black circles show the observations. The blue lines, dark shaded and light-shaded areas show the median, one-sigma (i.e. 68\%) and two-sigma (i.e. 95\%) of 5000 Monte Carlo samplings of the Random model, wherein GLEs are assumed to have equal occurrence probability at all times. The red lines and shaded areas show the same for the Phase model, wherein GLEs are a factor four more probable in the active phase than in the quiet phase. }\label{fig:phase}
 \end{figure}

We first test the significance of the apparent solar cycle phase trend in GLE occurrence. Specifically, we test the degree to which the observed difference in GLE occurrence probability in the quiet and active phases of the solar cycle can be explained purely by the random occurrence of a small number of events. Thus the Random model is our null hypothesis, as it assumes there is no underlying solar cycle trend. The Phase model, in which there is an underlying trend in the occurrence of GLEs with solar cycle phase, is the proposed hypothesis.

For each GLE intensity threshold (i.e. $\log (F) > 2, 4.3, 4.7, 5.5$ cm$^{-2}$), we produce a random time series of GLE occurrence consistent with the Random and Phase models, and containing the appropriate number of events (i.e. 67, 32, 15, and 8, respectively). For each model time series, we compute the average GLE occurrence probability in the quiet and active phase of the solar cycle. This is done 5,000 times to produce a Monte Carlo sampling of the average GLE occurrence probabilities in the quiet and active phases of the solar cycle. 

Figure \ref{fig:phase} shows the median, 1- and 2-sigma ranges (i.e. containing 68 and 95\% of the samples) of the Monte Carlo samples of the model values. In the quiet phase (Figure \ref{fig:phase}a), the Random model overestimates the observed occurrence probability for all GLE thresholds, while the Phase model is largely in agreement with the observations. Similarly, in the active phase (Figure \ref{fig:phase}b), the Random model underestimates the occurrence probability. These two trends are combined in Figure \ref{fig:phase}c, which shows the difference between the active and quiet phase occurrence probability. The Random model is centred on zero, as expected. The observed values are well outside of 95\% of the Random model values. Thus we can say that the Random model, in which there is no underlying solar cycle phase trend in GLE occurrence, is only consistent with the observations at a probability $p < 0.05$. Conversely, the Phase model describes the observations well.

\subsection{Solar Cycle Amplitude}
\label{sec:amp}

 \begin{figure}[!t]
 \centerline{\includegraphics[width=0.6\textwidth,clip=]{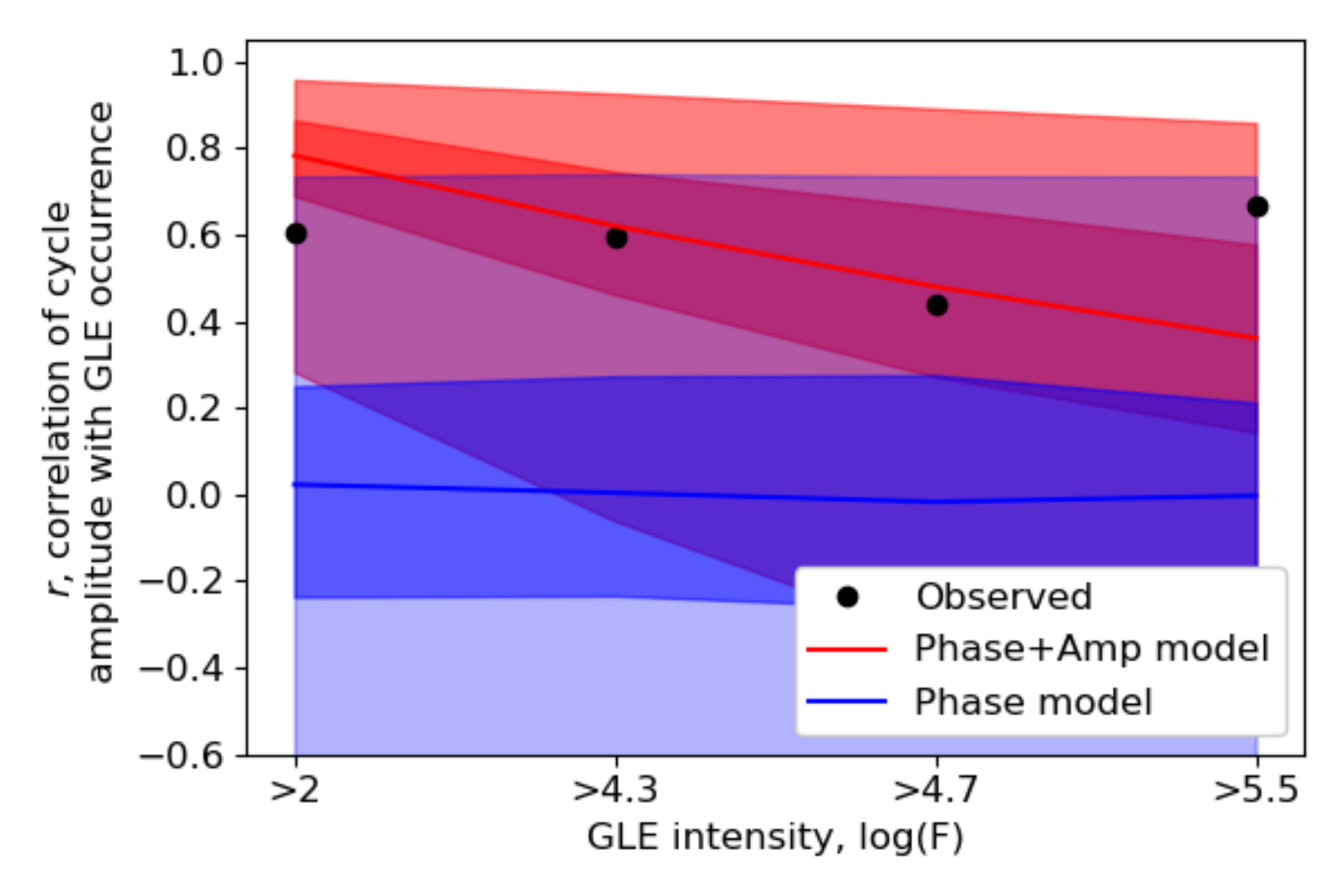}}
 \caption{The linear correlation, $r_L$, between solar cycle amplitude (as measured by the cycle-average sunspot number) and average GLE occurrence probability. Observations are shown by black circles. The blue lines, dark- and light-blue shaded areas show the median, one-sigma (i.e. 68\%) and two-sigma (i.e. 95\%) of 5000 Monte Carlo samplings of the Phase model, wherein GLEs are assumed to have equal occurrence probability in all cycles. The red lines and shaded areas show the same for the Phase+Amp model, wherein GLE occurrence probability increases linearly with cycle amplitude.}\label{fig:cyclemagnitude}
 \end{figure} 

The next test is to determine whether larger cycles produce more GLEs. For each GLE threshold, we compute the linear (or Pearson) correlation coefficient, $r_L$, between the cycle amplitude -- as characterised by $SSN_N$ -- and the average GLE occurrence probability over the cycle, $<p_{GLE}>_N$. For all thresholds, this correlation is based on only six data points (the six solar cycles in the dataset). However, if there is an underlying relation between $SSN_N$ and $<p_{GLE}>_N$, we might expect the correlation to drop with GLE threshold as $<p_{GLE}>_N$ becomes more poorly defined, being based on fewer events per cycle.

Figure \ref{fig:cyclemagnitude} shows the observed correlation between $SSN_N$ and $<p_{GLE}>_N$ as a function of GLE intensity. The correlation remains reasonably constant at around $r_L = 0.6$. At each threshold, $N = 6$, thus using a Student t-test there is a reasonable probability ($p = 0.2$) that the underlying correlation is zero and that this observed value occurred merely by chance. The Monte Carlo tests using the probability models reach a similar conclusion; while the  observations agree more closely with the Phase+Amp model than the null hypothesis of the Phase model (wherein there is no relation between solar cycle amplitude and GLE occurrence), the null hypothesis still has a large probability ($p> 0.05$) of describing the data.

\subsection{Solar Cycle Parity}
\label{sec:parity}

 \begin{figure}[!t]
 \centerline{\includegraphics[width=0.6\textwidth,clip=]{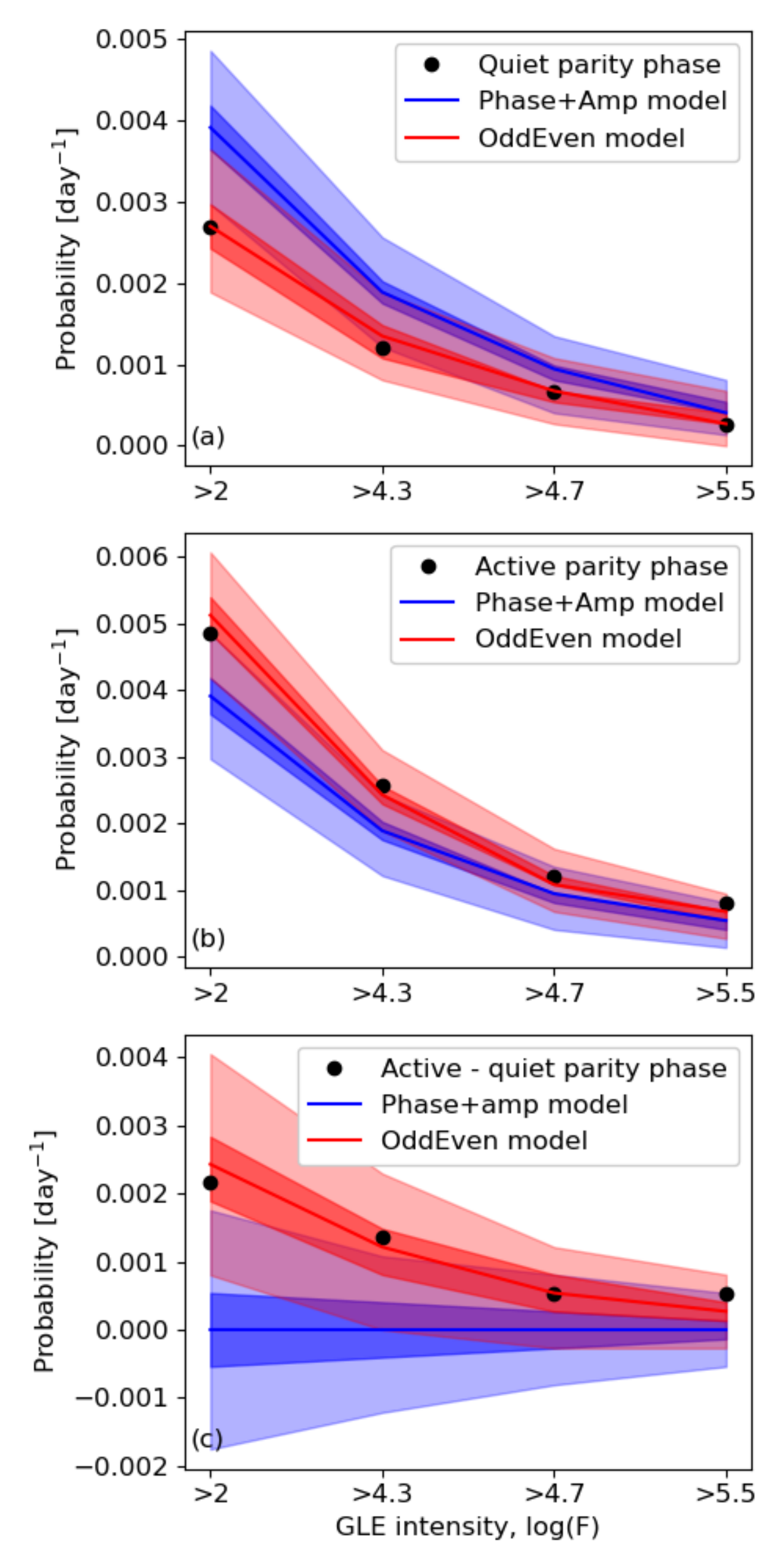}}
 \caption{GLE occurrence probability with increasing intensity threshold (a) in the quiet parity phases, i.e. early in even cycles, late in odd cycles, (b) in the active parity phases, i.e. late in even cycles, early in odd cycles, and (c) the difference between the active and quiet parity phases. Observations are shown in black. The blue line, dark shaded and light-shaded areas show the median, one-sigma (i.e. 68\%) and two-sigma (i.e. 95\%) of 5000 Monte Carlo samplings of the Phase+Amp model, wherein GLEs are assumed to have equal occurrence probability in the early and late active period. The red lines and shaded areas show the same for the OddEven model, wherein GLEs are a factor more probable in the early active phase in even cycles and in the late active phase in odd cycles. }\label{fig:oddeven}
 \end{figure}

Next we test the apparent difference in GLE occurrence during solar cycles of different parity (i.e. even- and odd-numbered solar cycles). The Phase+Amp model serves as the null hypothesis, as it contains no difference between odd and even cycles. The proposed hypothesis -- that there is an underlying preference for activity occurs earlier in even cycles and later in odd cycles -- is tested using the OddEven model. 

Figure \ref{fig:oddeven}a shows the average GLE occurrence probability in the combined quiet parity phases, i.e. late in the active phase of even cycles and early in the active phase of odd cycles. Figure \ref{fig:oddeven}a shows the same for the active parity phases (early in even cycles, later in odd cycles). The null hypothesis of no difference between odd and even cycles significantly overestimates GLE occurrence in the quiet parity phase at all GLE magnitudes, and systemically underestimates occurrence in the active phase (though cannot be ruled out at the 95\% confidence level). The Phase+Amp model, on the other hand, is in general agreement. Figure \ref{fig:oddeven}c shows the difference between these active and quiet parity phases. For the Phase+Amp model, the median value of the difference is zero, as expected, as there is no difference between odd and even cycles. It is clear that the observations are more consistent with the OddEven model. For three of the four GLE intensity thresholds considered, the observations are consistent with the null hypothesis with $p < 0.05$.

\subsection{GLE Waiting Times}

  \begin{figure}[!t]
 \centerline{\includegraphics[width=1\textwidth,clip=]{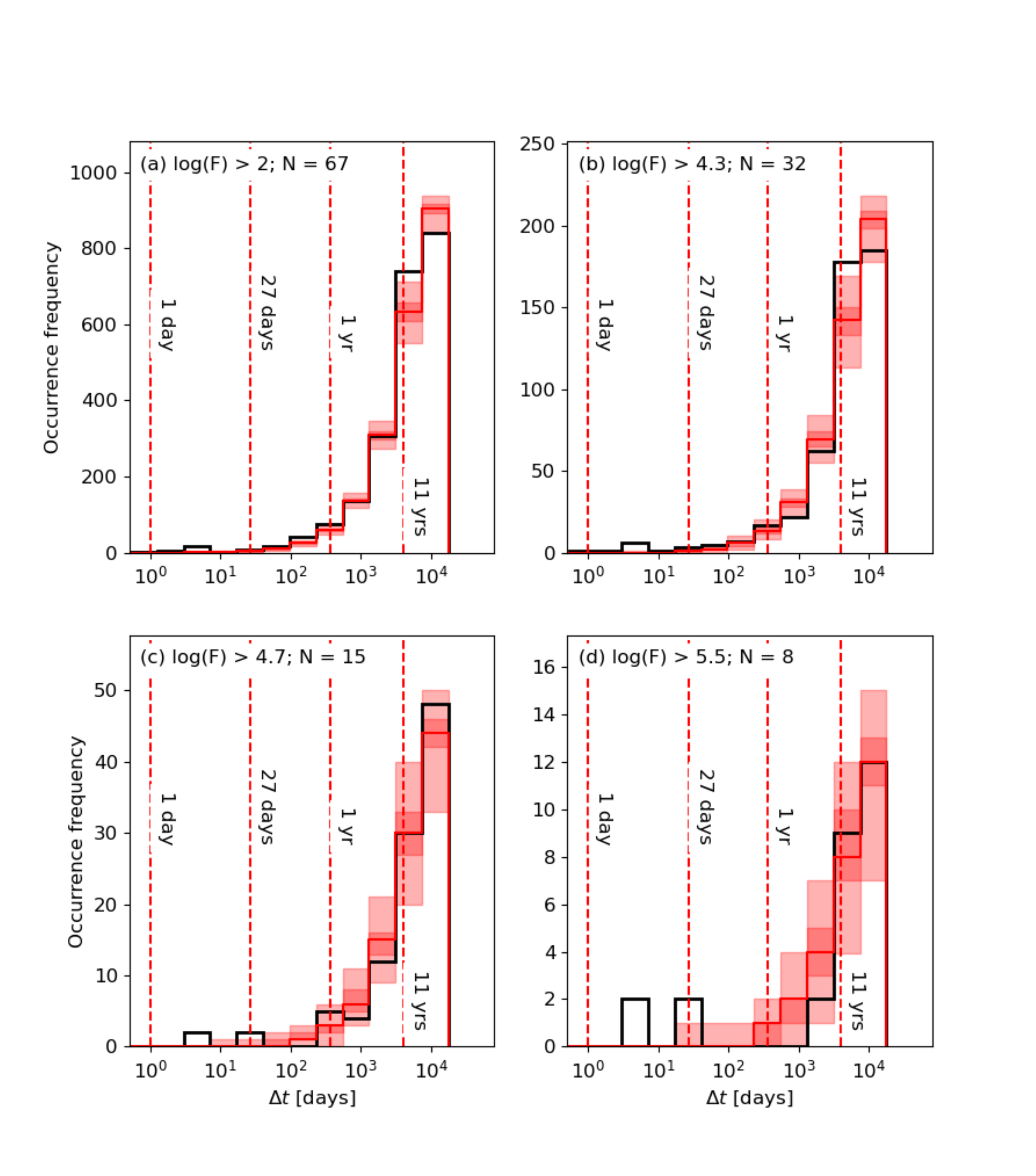}}
 \caption{Histograms of GLE waiting times for events defined by four different intensity thresholds. Note that the bins are equally spaced in log space. Red dashed vertical lines show a number of times of interest, for reference. Observations are shown in black. The red line, dark shaded and light-shaded areas show the median, one-sigma (i.e. 68\%) and two-sigma (i.e. 95\%) of Monte Carlo samples of the Random model. }\label{fig:waitingtimes}
 \end{figure}
 
   \begin{figure}[!t]
 \centerline{\includegraphics[width=1\textwidth,clip=]{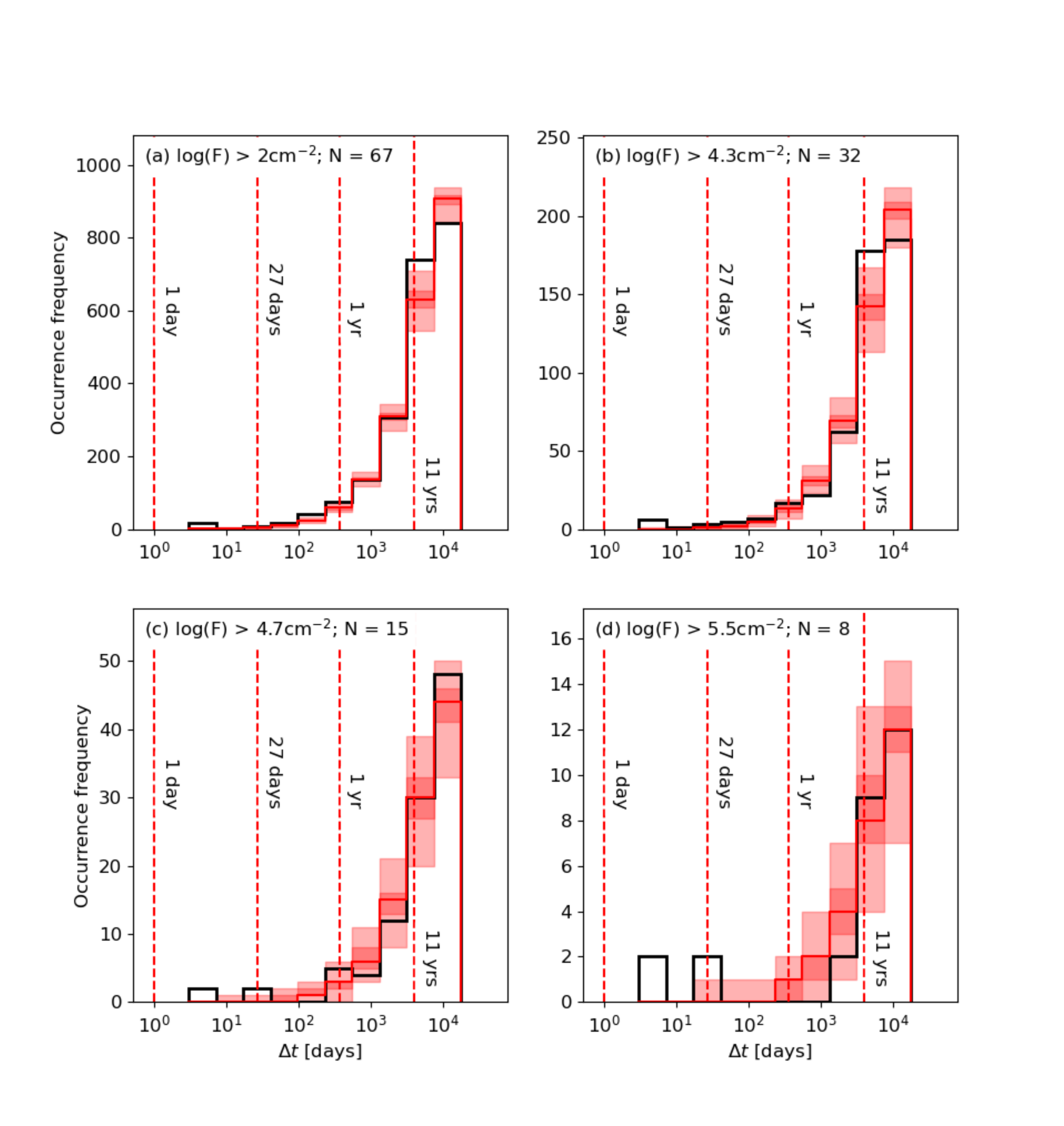}}
 \caption{Histograms of the separation times, $\Delta t$, between all possible GLE pairs for events defined by four different intensity thresholds. Note that the bins are equally spaced in log space. Red dashed vertical lines show a number of times of interest, for reference. Observations are shown in black. The red line, dark shaded and light-shaded areas show the median, one-sigma (i.e. 68\%) and two-sigma (i.e. 95\%) of Monte Carlo samples of the Random model. }\label{fig:dt}
 \end{figure}

Some of the proposed Miyake events can require enhanced $^{14}$C production over multiple consecutive years \citep{sakurai_prolonged_2020}. Given SEPs typically occur on the timescale of hours to days, multi-year $^{14}$C production requires clustering of extreme SEPs over a period of several months to years. Quasi-periodicities on these time scales have been reported for GLEs \citep{perez-peraza_prognosis_2015, marquez-adame_determination_2019}, though the statistics are dominated by the smaller magnitude events. We here examine the GLE record to assess the likelihood of such clustering of the most extreme events.

Figure \ref{fig:waitingtimes} shows the waiting time between consecutive GLEs for different GLE amplitudes. The red shaded areas show the 1- and 2-sigma ranges of 5000 Monte Carlo samples of the Random model, i.e. containing 68 and 95\% of the samples, respectively. Other models (Phase, Phase+Amp and OddEven) are not shown as they do not differ significantly from the Random model in that the events occur without correlation. At the lowest threshold shown in Figure \ref{fig:waitingtimes}a, there are 16 events that occur within 10 days of a previous GLE. The clustering of these events are inconsistent with the random occurrence of events, and none of the other models considered in this study would produce such short-term clustering. This effect is likely multiple GLEs produced by a single, long-lived active region. Memory on these short time scales has been reported for lower intensity SEPs observed in near-Earth space \citep{jiggens_time_2009}, though CME and flare waiting times appear to follow a time-dependant Poisson process \citep{wheatland_coronal_2003}. 

There is a large peak in events approximately 1 year after the previous GLE, but this is consistent with random occurrence of 67 events through the 60 year interval. This expected peak at approximately 1-year waiting time between consecutive GLEs persists out to $\log F > 4.7$ cm$^{-2}$. For the higher thresholds there is not strong evidence for clustering at the 1-2 year timescale, but only 8 events are available for analysis. As GLE magnitude increases, the 11-year peak becomes more prominent, with GLEs being separated by a whole solar cycle, though the statistical significance is low.

Next, we look at the separation time between each possible GLE pair, $\Delta t$, shown in Figure \ref{fig:dt}. In this dataset, $\Delta t$ can vary between the resolution of the data, at one day, and the length of the time series, approximately 60 years. As shown by comparison with the Monte Carlo sampling of the Random model, there is more clustering observed at very short times (less than a month) than would be expected. The solar cycle ordering of GLEs can be seen at lower event thresholds as a slight enhancement around $\Delta t = 11$ yrs. 

For both waiting time and $\Delta t$, we note some suggestion of a small peak around 27 days at all thresholds, suggesting some recurrent activity (though it could be associated with CMEs originating from the same active region, rather than necessarily from recurrent activity driven by corotating interaction regions).

\section{Discussion}
\label{sec:discussion}

In order to both better understand extreme space weather and to aid in the interpretation of `Miyake events' in the cosmogenic-isotope records, this study has investigated solar cycle trends in the solar energetic particle (SEP)  ground-level enhancement (GLE) record during the neutron monitor era. As there are only 67 events since 1956 (68 including the recent Solar Cycle 25 event), spanning only 6 solar cycles, it is necessary to carefully evaluate the probability that any apparent trends in GLE occurrence are not merely the result of random variations with a small sample size.

While such solar cycle trends were recently investigated for extreme geomagnetic storms, we show here that the one-to-one event association between GLEs and extreme geomagnetic storms is weak; most GLEs are not associated with a major geomagnetic storm and most major geomagnetic storms are not associated with a GLE. Thus solar cycle behaviour of storms should not be assumed a priori to apply to GLEs. The lack of a one-to-one event association is not surprising, as extreme storms require Earth-directed CMEs, whereas highest SEP fluence and hardest SEP spectra result from eruptions and shocks along the Earth-connected heliospheric magnetic flux tube \citep[e.g.][]{reames_solar_1995}. Owing to the Parker spiral configuration of the heliospheric magnetic field \citep{parker_dynamics_1958}, this is nominally between 30 and 60 degrees West of Earth for coronal shocks. As the most SEP-productive shocks are fast and wide \citep{gopalswamy_coronal_2008}, the same shock and CME can be both directed along the Earth-connected HMF in the corona and encountered at Earth. However, for the most extreme storms it is likely the `nose' of the shock (and thus typically the centre of the CME) that arrives at Earth. Similarly, for the most extreme SEPs, the nose of the shock likely threads the Earth-connected Parker spiral.  Thus, the CMEs producing the highest SEP fluences at Earth are not Earth directed, and the most geoeffective CMEs are not directed along the Parker spiral-connected West limb of the Sun. 

Despite the lack of association between events driving extreme storms and GLEs, the statistical behaviour of storm and GLE occurrence is remarkably similar. Thus GLEs act as independent sources of evidence of underlying solar-cycle trends in extreme solar activity reported by \cite{owens_extreme_2021}. As with extreme geomagnetic storms, there is a clear solar cycle trend in GLE occurrence, with around a factor four increase in occurrence probability in an active period centred on solar maximum, compared with a quiet period centred on solar minimum. This trend is weaker in amplitude than reported for geomagnetic storms \citep{owens_extreme_2021}. Perhaps more surprisingly, GLEs also exhibit the 22-year trend for preferential occurrence earlier in even-numbered solar cycles and later in odd-numbered cycles. The most likely explanation for this trend is differing energetic particle drift patterns during opposing solar magnetic field polarities, which have long been known to affect GCR propagation through the heliosphere \citep{usoskin_history_2017} and have recently been shown to have a significant effect on SEPs and GLEs \citep{waterfall_modelling_2022}. The relation between solar cycle amplitude and GLE occurrence cannot be conclusively measured, possibly owing to the few (six) solar cycles available for study. However, while the null hypothesis -- of no relation -- cannot be discounted, we note that: (a) the Phase+Amp model does describe the data better than the null hypothesis, and (b) extreme geomagnetic storm occurrence does exhibit a solar cycle amplitude trend \citep{owens_extreme_2021}, and GLEs follow the other two solar cycle trends in the same way as storms. Thus on balance, it seem more likely than not that larger solar cycles produce more GLEs.

There are a number of implications of these results for Miyake events (i.e., the spikes in the cosmogenic-isotope records). If Miyake events are more extreme GLEs, which are themselves extreme SEPs, then we might expect the same underlying patterns of occurrence. To assess if Miyake event follow the solar cycle phase trend requires multiple events, as it is inherently a statistical relation, and the probability difference between active and quiet phases is only around a factor four. It would also require precise determination of both the timing of the cosmogenic-isotope production enhancement and of the associated phase of the solar cycle. This would need to be within around 0.2 solar cycle phase, which is approximately 2 years. This may be difficult owing to the low signal-to-noise of the solar cycle in the $^{14}$C record, which often necessitates additional smoothing to identify \citep{brehm_eleven-year_2021}. On the other hand, $^{10}$Be in ice cores may have dating uncertainties of several years \citep[e.g.,][]{sukhodolov_atmospheric_2017}. The presence of a Miyake event further disturbs the underlying solar cycle signal, meaning phase must often be extrapolated from previous and subsequent cycles \citep{usoskin_solar_2021}, adding uncertainty to the estimate. Even more difficult to assess from the $^{14}$C data is the change of behaviour between odd- and even-numbered solar cycles. The parity of the solar cycle cannot simply be tracked back from modern times due to breaks in cycle numbering through grand minima like the Maunder minimum. However, it has been shown that solar cycle parity can be reconstructed from the shape of the variation in $^{10}$Be data \citep{owens_heliospheric_2015}. Using similar measures it may be possible to infer cycle parity from $^{14}$C data and determine if the early/late behaviour is present for Miyake events.

Extreme SEPs are -- by definition -- rare. Therefore a physical explanation of $^{14}$C spikes that requires multiple extreme SEPs in a relatively short space of time becomes less probable, unless the occurrence of such events is expected to preferentially cluster in time over similar time scales. At lower GLE thresholds, a significant fraction of GLEs do occur within 1 year of previous events, though only at a rate consistent with random occurrence. As higher event thresholds are considered, times between GLEs shift to a more bimodal distribution of a few days, owing to multiple events produced by a single active region, and around 11 years, owing to the solar-cycle ordering of occurrence. These preferential time scales are too short and too long, respectively, to provide the required extended $^{14}$C production spike. Thus such Miyake events that can require multiple years of enhanced $^{14}$C production \citep[e.g. circa 660 BCE;][]{sakurai_prolonged_2020} are less easy to explain in terms of solar-generated energetic particle events, if we think of them as the extreme event tail of the distribution to which the observed GLE events belong.

%
%

%

%

%
\begin{acks}
We have also benefited from sunspot data provided by Royal Observatory of Belgium SILSO and GLE data by the GLE database.
\end{acks}

\begin{fundinginformation}
This work was part-funded by Science and Technology Facilities Council (STFC) grant numbers ST/R000921/1 and ST/V000497/1, and Natural Environment Research Council (NERC) grant numbers NE/S010033/1 and NE/P016928/1. I.U. acknowledges the Academy of Finland (projects ESPERA No. 321882). E.A. acknowledges support from the Academy of Finland (Postdoctoral Researcher Grant 322455), and the Finnish Centre of Excellence in Research of Sustainable Space (Academy of Finland grant number 312390).
\end{fundinginformation}

 \begin{dataavailability}
Sunspot data are provided by the Royal Observatory of Belgium SILSO and available from \url{www.sidc.be/silso/DATA/SN_m_tot_V2.0.csv}. The GLE database can be accessed here: \url{https://gle.oulu.fi/}. The  aa$_\mathrm{H}$-geomagnetic index data are available from \url{www.swsc-journal.org/articles/swsc/olm/2018/01/swsc180022/swsc180022-2-olm.txt}.

All analysis and visualisation code is packaged with all required OMNI data here: \url{https://github.com/University-of-Reading-Space-Science/ExtremeEvents}. 
 \end{dataavailability}

 \begin{ethics}

\textbf{Disclosure of Potential Conflict of Interest}: We declare we have no conflicts of interest.

 \end{ethics}

%
%

%
%
%
%

\end{article}
\end{document}